\documentstyle[titlepage,12pt]{article}
\begin{document}

\begin{titlepage}
\renewcommand{\thefootnote}{\fnsymbol{footnote}}
\hfill{ITP-SB-92-63}

\vskip 2cm

\begin{center}
{\large\bf The One-Loop Five-Graviton Scattering Amplitude and
Its Low-Energy Limit\(\footnote[2]{{\rm Work supported in part by
National Science Foundation Grant
PHY 90-08936.}}\)}

\vskip 1.5cm

J. Lee Montag

\vskip .5cm

Institute for Theoretical Physics \\
State University of New York at Stony Brook \\
Stony Brook, NY 11794-3840

\vskip 1.5cm

\bf ABSTRACT

\end{center}

\vskip .1cm

A covariant path integral calculation of the even spin structure
contribution to the one-loop
N-graviton scattering amplitude in the type-II
superstring theory is presented. The apparent divergence
of the $N=5$ amplitude is resolved by separating it into twelve
independent terms corresponding to different orders of inserting the
graviton vertex operators. Each term is well defined in an appropriate
kinematic region and can be analytically continued to physical regions
where it develops branch cuts required by unitarity. The zero-slope
limit of the $N=5$ amplitude is performed, and the Feynman diagram
content of the low-energy field theory is examined. Both one-particle
irreducible (1PI) and one-particle redicible (1PR) graphs with
massless internal states are generated in this limit. One set of 1PI
graphs has the same divergent dependence on the cut-off as that found
in the four-graviton case, and it is proved that such graphs exist
for all~$N$. The 1PR graphs are contributed by the poles in the
world-sheet chiral Green functions.

\newpage
\end{titlepage}

\newcommand{\beq}{\begin{equation}}
\newcommand{\eeq}{\end{equation}}
\newcommand{\beqa}{\begin{eqnarray}}
\newcommand{\eeqa}{\end{eqnarray}}
\newcommand{\al}{\alpha^{\prime}}
\newcommand{\ov}{\overline}
\newcommand{\half}{\frac{1}{2}}

\section{Introduction}

We present an explicit calculation of the one-loop
five-graviton amplitude in the critical NSR \mbox{type-II}
string theory using path integral methods,
resolve its apparent divergence,
and study its low-energy field theory content.\footnote{We use the
word graviton to include all of the
massless vector bosons of the theory (graviton, dilaton, and
anti-symmetric tensor field).}
In a previous paper~\cite{me}, we resolved
the divergence of the one-loop four-graviton
scattering amplitude in the \mbox{type-II} superstring theory
by separating the amplitude into terms corresponding to
different orders of inserting the graviton vertex operators
on the world-sheet. The
scalar factor for each term contains the proper-time representation
of a box type Feynman diagram with the internal lines summed over
the mass levels of the superstring spectrum. This decomposition of
the amplitude, however, does not directly give terms representing
diagrams with single-particle intermediate states which are required
by unitarity. In the case of the four-graviton
amplitude such terms result from duality by summing over an infinite
number of mass levels on a given internal line in the box diagram.
No massless single poles appear in the limit of the sum, however,
and this is consistent with the fact that the two- and three-graviton
amplitudes vanish on the torus.

The one-loop five-graviton amplitude does contain a massless
pole in two to three particle channels in the limit when two vertex
operators come close together on the world-sheet and the sum of their
respective momenta goes on the graviton mass shell.
The residue factorizes into
a tree level three-point amplitude times a one-loop four-point amplitude.
Therefore, the integrand of the one-loop five-point function must contain
more than just a simple extension of the exponential dependence on the
bosonic world-sheet Green function found in the one-loop four-point
function. One finds no massless pole, however, in the limit where
three vertex operators come close together, and the sum of their
respective momenta goes on the graviton mass shell.
This is consistent once again
with the fact that the three-graviton amplitude vanishes on the torus.

A calculation of the one-loop five-graviton scattering
amplitude using operator methods was
previously presented in~\cite{frampton}. The path integral
approach that we follow, however,
makes the modular invariance and the field
theory limit more straightforward.
In contrast to the four-graviton amplitude where the massive single
particle poles arise only from summing over box diagrams
(i.e. from duality), we find that
the one-particle reducible (1PR) diagrams
with a massless single pole appear directly in the zero-slope limit
of the five-graviton amplitude.
These massless single-particle poles are absent in
the four-graviton amplitude because the only dependence
of the integrand on the bosonic world-sheet coordinates of the
vertex operators comes from the exponential of the bosonic
world-sheet Green function.
This simple structure results from the Grassman
integration over fermionic world-sheet coordinates and a Riemann
identity for theta functions.
We find that in the case of the five-graviton amplitude no
such simplification exists, and the dependence of the integrand on
chiral Green functions cannot be avoided.

In addition to the 1PR diagrams, two sets of
one-particle irreducible (1PI) graphs with
massless internal states also occur in the zero-slope limit of the
five-graviton amplitude. In each set, all possible distinct
channels contribute. These distinct channels arise from dividing the
integration region of the imaginary part of the coordinates for the
vertex operators into their different possible orderings in exact
analogy to the four-graviton amplitude~\cite{me}.
Once again, each
ordering must be independently analytically continued in the external
momenta to the physical region for scattering in order that a finite
answer for the amplitude be obtained without violating momentum
conservation or the graviton mass shell condition. A straightforward
transformation from the world-sheet coordinates to Feynman parameters
can be found in the zero-slope limit, and a
convenient set of momentum invariants
exists for each channel which makes the identification of the
two-particle cuts trivial.
We argue that higher order terms in the low-energy
expansion correspond to Feynman diagrams with
massive internal states from the superstring spectrum.

One set of 1PI graphs has the expected logarithmic divergence in
the cut-off~$\al$~\cite{polchinski}
for a scalar five-point amplitude in ten dimensions
and contains powers of
the Feynman parameters in the integrand corresponding to derivative
couplings in the low-energy field theory.
We also find another set of 1PI graphs appearing in the
zero-slope limit which has
the same degree of divergence as that found in the
four-point amplitude. The term that generates these
graphs has exactly the same form as in the
four-graviton case (i.e. a kinematic factor times a scalar
integral), but with the product over world-sheet
bosonic Green functions extended to incude the extra particle.
Therefore, in the zero-slope limit one obtains 1PI graphs
with cubic interactions but with the wrong power of \mbox{Im\,$\tau$}
in the denominator of the integrand
for a scalar five-point amplitude in ten dimensions.
One can prove this result in string theory
from modular invariance. The one-loop four- and five-graviton
amplitudes receive contributions only from the three even spin
structures on the torus. The term generating
these particular 1PI graphs in the zero-slope limit
comes from the even spin structure part of the amplitude.
Since the integrand of this term
contains the same modular invariant function no matter how many
external gravitons are inserted, its dependence
on the proper-time variable Im\,$\tau$ is fixed by the
requirement that the entire integral be modular invariant.
It may then be shown that these 1PI graphs
will have the same linear divergence in
the cut-off to all orders in the number
of external gravitons. Therefore, modular invariance from
string theory tells us that inserting additional external
gravitons will not change the degree of divergence of this set of 1PI
Feynman graphs in the low-energy field theory.

In sect.~2 we present an explicit outline of the
calculation of the even spin structure contribution to the genus-one
N-graviton scattering amplitude in the \mbox{type-II}
superstring theory using the NSR path integral formulation.
Although this calculation was given
by D'Hoker and Phong~\cite{D'Hoker} in the superspace approach,
the component formulation presented here makes the identification
of the low-energy structure more straightforward.
In sect.~3 we show how the four- and
five-graviton amplitudes may be obtained
from the general N-graviton result by performing the necessary
Grassman integrations and verify that they are modular invariant.
In sect.~4 we examine the Feynman diagram content of the
zero-slope limit of the five-graviton amplitude and discuss the
unusual ultra-violet divergence in the cut-off contained in the 1PI part.
In sect.~5 we present some concluding remarks.

\section{The N-graviton amplitude}

We begin the calculation of the even spin structure contribution to
the one-loop N-graviton scattering amplitude by quoting the result
for the partition function~\cite{D'Hoker} on the torus with fixed
even spin structure $[\nu_{a},\nu_{b}]$ for the left movers
and $[\ov{\nu}_{a},\ov{\nu}_{b}]$ for
the right movers.\footnote{$[\nu_{a},\nu_{b}]$ can
take on values $[\frac{1}{2},0]$, $[0,\frac{1}{2}]$, and $[0,0]$
for even spin structure. We denote these by the subscript
\mbox{ab = 10, 01, and 00} respectively.}
We suppress spacetime indices for the fields when unambiguous.
\beqa
{\cal Z}_{\nu \ov{\nu}}
&=& \int d\mu
\int [D\psi] [D\overline{\psi}] [Dx]
\,\, e^{-S}   \nonumber\\
S
&=& \frac{1}{2 \pi \al} \int id^{2}z
\left( \partial x \overline{\partial} x \, + \,
\frac{1}{2} \psi \overline{\partial} \psi \, - \,
\frac{1}{2} \overline{\psi} \partial \overline{\psi}
\right)  \, ,
\eeqa
where the gauged fixed measure for the integral over inequivalent
world-sheet metrics on the torus is given by
\beq
d\mu = \frac{d^{2}\tau}{\tau_{2}} \left| \eta (\tau) \right|^{4}
\left| \frac{\Theta_{ab}(0|\tau)}{\eta (\tau)} \right|^{-2} \, .
\eeq
The complex modular parameter for the torus is
$\tau = \tau_{1} + i\tau_{2}$, and the Dedekind eta function is given by
$\eta (\tau)=\exp{[\frac{i \pi \tau}{12}]} \prod_{n=1}^{\infty}
\left( 1-e^{2i \pi n\tau} \right)$. $\Theta_{ab}(z|\tau)$
is the Jacobi theta function with characteristics
$[\nu_{a},\nu_{b}]$ (see appendix).

We obtain the
even spin structure contribution to the N-graviton scattering
amplitude by inserting N vertex operators of the
following form~\cite{D'Hoker}.
\beqa
V(z, \ov{z} ; \theta , \ov{\theta} ; p)
&=& \kappa \epsilon_{\mu \nu} D_{+}\Phi^{\mu} D_{-}\Phi^{\nu}
\,\, e^{i p \cdot \Phi} \nonumber\\
\Phi^{\mu}(z,\overline{z};\theta,\overline{\theta})
&=& x^{\mu} + \theta \overline{\psi}^{\mu}
- \overline{\theta} \psi^{\mu} \, ,
\eeqa
where the superderivatives in the superconformal gauge are given by
\beqa
D_{+} &=& \frac{\partial}{\partial \overline{\theta}}
- 2\overline{\theta} \partial  \nonumber\\
D_{-} &=& \frac{\partial}{\partial \theta}
+ 2\theta \overline{\partial} \, ,
\eeqa
and $\kappa$ is the gravitational coupling constant with
dimension $[L]^{4}$ for $d=10$.
The world-sheet matter integrals are easily
performed after introducing Grassman
sources for the superderivative pieces. We therefore take the
expectation value of the following operator,
\beq
{\cal O}(\eta,\overline{\eta})
= \exp{ \sum_{i=1}^{N} \left[ i p_{i} \cdot \Phi_{i}
+ i \overline{\eta}_{i} \cdot D_{+} \Phi_{i}
+ i \eta_{i} \cdot D_{-} \Phi_{i} \right] } \, .
\eeq
The result can be written in terms of the world-sheet Green
functions $G_{ab}$ and $S_{ab}$ defined by
\beqa
\overline{\partial} \partial G_{ab}(z,\overline{z})
&=& \delta^{(2)}(z,\overline{z})
+ \frac{1}{2i \tau_{2}} \delta_{a,1} \delta_{b,1} \nonumber\\
S_{ab}(z)
&=&  \partial G_{ab}(z,\overline{z}) \, .
\eeqa
The particular Green functions useful for our purposes are
\beqa
G_{11}(z,\overline{z})
&=& -\frac{1}{2i \pi} \log{ \left|
e^{-\pi y^{2}/ \tau_{2}}
\frac{\Theta_{11}(z| \tau)}{\eta(\tau)} \right|^{2} } \nonumber\\
S_{ab}(z)
&=& -\frac{1}{2i \pi} \frac{\Theta_{ab}(z| \tau) \Theta_{11}^{'}(0| \tau)}
{\Theta_{ab}(0| \tau) \Theta_{11}(z| \tau)} ,
\,\,\,\,\,\, ab \neq 11 \, .
\label{Gdef}
\eeqa

The N-graviton scattering amplitude for fixed even spin
structure in both the left and
right moving sectors is given by\footnote{The primes
on the sums indicate that the $i=j$ terms
are excluded as a consequence of normal ordering. The superscripts
on the world-sheet Green functions label their \mbox{z-dependence}
$G_{11}^{ij}=G_{11}(z_{i}-z_{j})$.}
\beqa
\Gamma_{\nu \ov{\nu}}
&=& \frac{\kappa^{N}}{(\al)^{7-N}}
\int_{{\cal F}} \frac{d^{2} \tau}{(\tau_{2})^{6}}
\int \left[ \prod_{i=1}^{N}
\epsilon_{i}^{\mu_{i} \nu_{i}}
d\ov{\eta}_{i\mu_{i}} d\eta_{i\nu_{i}}
d^{2}\theta_{i} d^{2}z_{i} \right] \nonumber\\
& & \times
\exp{ \left\{ -\frac{i\pi \al}{2}
\sum_{ij}{}^{'} p_{i} \cdot p_{j}
\left[ G_{11}^{ij} + \frac{1}{4i\tau_{2}} (z_{ij} -\ov{z}_{ij})^{2}
\right] \right\} } \nonumber\\
& & \times
\left| \frac{\Theta_{ab}(0| \tau)}{[\eta(\tau)]^{3}} \right|^{8}
\exp{ \left[ K_{ab}(\overline{\eta},\overline{\theta})
\,+\, \ov{K}_{\ov{a}\ov{b}}(\eta,\theta) \right]} \nonumber\\
& & \times \,\, {\rm exp}  \left\{
\frac{\pi \al}{\tau_{2}} \sum_{ij}^{}{'} \left[
p_{i} \cdot \ov{\eta}_{j} \ov{\theta}_{j} (z_{ij}-\ov{z}_{ij})
+ p_{i} \cdot \eta_{j} \theta_{j} (z_{ij}-\ov{z}_{ij})
\right. \right. \nonumber\\
& & + \left. \left.
\frac{1}{8} p_{i} \cdot p_{j} (z_{ij} -\ov{z}_{ij})^{2}
+ 2\ov{\eta}_{i} \cdot \eta_{j} \ov{\theta}_{i} \theta_{j}
+ \ov{\eta}_{i} \cdot \ov{\eta}_{j} \ov{\theta}_{i} \ov{\theta}_{j}
+ \eta_{i} \cdot \eta_{j} \theta_{i} \theta_{j}
\right] \right\}   ,
\eeqa
where the chiral piece $K_{ab}(\ov{\eta}, \ov{\theta})$
depending holomorphically on $z$ and $\tau$ is
\beqa
K_{ab}(\overline{\eta},\overline{\theta})
&=& -i\pi \al \sum_{ij}{}^{'} \left\{
p_{i} \cdot p_{j}
\overline{\theta}_{i} \overline{\theta}_{j} S_{ab}^{ij}
\,+\, \overline{\eta}_{i} \cdot \overline{\eta}_{j} S_{ab}^{ij}
\,-\, 2p_{i} \cdot \overline{\eta}_{j} \overline{\theta}_{i} S_{ab}^{ij}
\right. \nonumber\\
& &
-\, 2p_{i} \cdot \overline{\eta}_{j} \overline{\theta}_{j}
\left[ (\partial_{j}G_{11}^{ij})
- \frac{1}{2i\tau_{2}} (z_{ij}-\ov{z}_{ij}) \right] \nonumber\\
& & \left.
-\, 2\overline{\eta}_{i} \cdot \overline{\eta}_{j}
\overline{\theta}_{i} \overline{\theta}_{j}
\left[ (\partial_{i} \partial_{j} G_{11}^{ij})
- \frac{1}{2i\tau_{2}} \right]
\right\} \, .
\eeqa

All terms proportional to delta functions have been dropped in
the above expression since they do not contribute to the integral
for values of the external momenta away from intermediate particle
poles. The complex modular parameter
$(\tau = \tau_{1} + i\tau_{2})$ is integrated over the fundamental
domain,
\beq
-\frac{1}{2} \leq \tau_{1} \leq \frac{1}{2} \, , \;\;\;
\tau_{2} > 0 \, , \;\;\; {\rm and} \;\;\; |\tau| \geq 1 \, ,
\eeq
and the coordinates of the vertex operators $(z_{i} = x_{i} + iy_{i})$
are integrated over the torus,
\beq
\frac{\tau_{1}}{\tau_{2}} y_{i} \leq x_{i} \leq
\frac{\tau_{1}}{\tau_{2}} y_{i} + 1 \, , \;\;\; {\rm and} \;\;\;
0 \leq y_{i} \leq \tau_{2} \, .
\eeq

Since the spin structures must be summed independently for the left
and right movers, we split the mixed chirality piece by introducing
the loop momenta as follows~\cite{D'Hoker}.
First, however, we include the finite $i=j$
terms in the sum which amounts to a finite renormalization of the
vertex. We can then re-write the mixed chirality piece as
\beq
\exp{ \left\{ -\frac{\pi \al}{\tau_{2}} \left[
\sum_{i} \left( \ov{\eta}_{i}^{\mu} \ov{\theta}_{i}
+ \eta_{i}^{\mu} \theta_{i} - \frac{1}{2}
p_{i}^{\mu} (z_{i} - \ov{z}_{i})
\right) \right]^{2} \right\} } .
\label{mch}
\eeq
We now split this term into a product of holomorphic and
anti-holomorphic factors with the following identity.
\beqa
1 &=&
\left( \frac{\tau_{2}}{4\al} \right)^{5}
\int \left[ \prod_{\mu} dq^{\mu} \right] \nonumber\\
& & \times \exp{ \left\{ -\frac{\pi \tau_{2}}{4\al}
\left[ q^{\mu} \,-\, \frac{2i\al}{\tau_{2}}
\sum_{i} \left( \ov{\eta}_{i}^{\mu} \ov{\theta}_{i}
+ \eta_{i}^{\mu} \theta_{i}
- \frac{1}{2} p_{i}^{\mu} (z_{i} -\ov{z}_{i}) \right)
\right]^{2} \right\} } .
\nonumber\\
\eeqa
This allows us to write~(\ref{mch}) as
\beqa
\left( \frac{\tau_{2}}{4\al} \right)^{5}
\int \left[ \prod_{\mu} dq^{\mu} \right]
\left| \exp{ \left[ \frac{i\pi \tau}{8\al} q \cdot q
+ i\pi \sum_{i} \left( q \cdot \overline{\eta}_{i} \overline{\theta}_{i}
- \frac{1}{2} q \cdot p_{i} z_{i} \right) \right] } \right|^{2} .
\eeqa

The result for the even spin structure contribution to the
N-graviton amplitude then factorizes into holomorphic and
anti-holomorphic pieces, and each is summed over even spin structures
with phases determined by the requirement of modular
invariance. We quote here the final result written in component form.
\beqa
\Gamma_{N}
&=& \frac{\kappa^{N}}{2^{10} (\al)^{12-N}}
\int_{{\cal F}} \frac{d^{2} \tau}{\tau_{2}}
\int \left[ \prod_{i=1}^{N}
\epsilon_{i}^{\mu_{i} \nu_{i}}
d\ov{\eta}_{i\mu_{i}} d\eta_{i\nu_{i}}
d^{2}\theta_{i} d^{2}z_{i} \right] \nonumber\\
& & \times
\left| \exp{ \left[ -\frac{i\pi \al}{2}
\sum_{ij}{}^{'} p_{i} \cdot p_{j} \hat{G}_{11}^{ij} \right] }
\right|^{2} \nonumber\\
& & \times
\int \left[ \prod_{\mu} dq^{\mu} \right]
\left| \exp{ \left( \frac{i\pi \tau}{8\al} q \cdot q
- \frac{i\pi}{2} \sum_{i} q \cdot p_{i} z_{i} \right) } \right|^{2}
\nonumber\\
& & \times  \left| \sum_{a \times b = 0} (-)^{a+b}
\left\{ \frac{\Theta_{ab}(0| \tau)}{[\eta(\tau)]^{3}} \right\}^{4}
\exp{ \left[ K_{ab} (\overline{\eta},\overline{\theta})
+ i\pi \sum_{i} q \cdot \ov{\eta}_{i} \ov{\theta}_{i}
\right] } \right|^{2} ,
\label{amp}
\eeqa
where we define $\hat{G}_{11}^{ij}$ to be the holomorphic
part of
$G_{11}^{ij} + \frac{1}{4i\tau_{2}} (z_{ij} -\ov{z}_{ij})^{2}$.

We have factorized the integrand of eq.~(\ref{amp})
into a piece depending holomorphically on $z$ and $\tau$ times
a piece depending anti-holomorphically on these variables.
Since some of these factors do not depend on either the
spin structure or the chiral Grassman sources,
for purposes of calculation it
is simpler to complete the square in the $q^{\mu}$ integral and
then shift $q^{\mu} \rightarrow q^{\mu} - \frac{i\al}{\tau_{2}}
\sum_{i} p^{\mu}_{i} (z_{i} - \ov{z}_{i})$. Using momentum
conservation we then define
\beq
\tilde{K}_{ab}(\ov{\eta}, \ov{\theta}; q)
= K_{ab}(\ov{\eta}, \ov{\theta}) + i\pi \sum_{j}
\left[ q - \frac{i\al}{\tau_{2}} \sum_{i}
p_{i} (z_{ij} - \ov{z}_{ij}) \right]
\cdot \ov{\eta}_{j} \ov{\theta}_{j} \, ,
\eeq
to write
\beqa
\Gamma_{N}
&=& \frac{\kappa^{N}}{2^{10} (\al)^{12-N}}
\int_{{\cal F}} \frac{d^{2} \tau}{\tau_{2}}
\int \left[ \prod_{i=1}^{N}
\epsilon_{i}^{\mu_{i} \nu_{i}}
d\ov{\eta}_{i\mu_{i}} d\eta_{i\nu_{i}}
d^{2}\theta_{i} d^{2}z_{i} \right] \nonumber\\
& & \times \exp{ \left[ -\frac{i\pi \al}{2}
\sum_{ij}{}^{'} p_{i} \cdot p_{j} G_{11}^{ij} \right]}
\int \left[ \prod_{\mu} dq^{\mu} \right]
\exp{ \left[ -\frac{\pi \tau_{2}}{4\al} q^{2} \right] } \nonumber\\
& & \times
\left| \sum_{a \times b = 0} (-)^{a+b}
\left\{ \frac{\Theta_{ab}(0| \tau)}{[\eta(\tau)]^{3}} \right\}^{4}
\exp{ \left[ \tilde{K}_{ab} (\overline{\eta},\overline{\theta};q)
\right] } \right|^{2}  .
\label{amp1}
\eeqa

\section{The five-graviton amplitude}

Before looking at the five-graviton amplitude,
we examine the case for $N<5$ in order to make clear
the different structure that appears at $N=5$. Instead
of expanding the exponential in eq.~(\ref{amp1}) and then
performing the Grassman integration, it is somewhat easier
in practice to
use the fact that this process is equivalent to Grassman
differentiation. Some useful expressions are
\beqa
\exp{ \left[ \tilde{K}_{ab} (\overline{\eta}=0,\overline{\theta};q) \right]}
&=& \prod_{1\le i < j \le N}
\left[ 1 \,-\, 2i\pi \al p_{i} \cdot p_{j}
\overline{\theta}_{i} \overline{\theta}_{j}
S_{ab}^{ij} \right] \, , \nonumber\\
\frac{\partial}{\partial \ov{\eta}_{j}^{\mu_{j}}}
\tilde{K}_{ab} (\overline{\eta}=0,\overline{\theta};q)
&=& 2i\pi \al \sum_{i \neq j}  p_{i\mu_{j}} \left[
\overline{\theta}_{i} S_{ab}^{ij}
\,+\, \overline{\theta}_{j}
\left( \partial_{j} G_{11}^{ij} \right) \right]
\,+\, i\pi q_{\mu_{j}} \ov{\theta}_{j}   \nonumber\\
\frac{\partial}{\partial \ov{\eta}_{i}^{\mu_{i}}}
\frac{\partial}{\partial \ov{\eta}_{j}^{\mu_{j}}}
\tilde{K}_{ab} (\ov{\eta}=0,\ov{\theta};q)
&=& 2i\pi \al g_{\mu_{i} \mu_{j}}
\left[ S_{ab}^{ij} \,-\, 2\ov{\theta}_{i} \ov{\theta}_{j}
\left( \partial_{i} \partial_{j} G_{11}^{ij}
- \frac{1}{2i\tau_{2}} \right) \right] \, .
\label{eta}
\eeqa
These equations allow one to differentiate using the chain rule
and to drop all those terms which would have cancelled anyway.

The contribution for $N=0$ vanishes trivially due to a Riemann
identity for theta functions (see appendix). In fact, for $N \leq 3$
the only contribution allowed by the Grassman integration over
the $\ov{\theta}_{i}$ consists of terms with products of either no
world-sheet fermion propagators $S_{ab}^{ij}$, two propagators, or
three propagators. In each of these cases the contributions
can be shown to vanish as a result of the same Riemann
identity~\cite{Namazie}.
When $N=4$, however, the Riemann identity gives instead of zero
\beq
\sum_{a \times b = 0}(-)^{a+b}
\left\{ \frac{\Theta_{ab}(0| \tau)}{[\eta (\tau)]^{3}} \right\}^{4}
S_{ab}^{i_{1}i_{2}}S_{ab}^{i_{2}i_{3}}
S_{ab}^{i_{3}i_{4}}S_{ab}^{i_{4}i_{1}}
= -1 \, ,
\label{riem}
\eeq
and eliminates the dependence of the integrand on the fermionic
Green function~$S_{ab}^{ij}$.

Since the above sum over spin structures is independent of
the world-sheet coordinates $z_{i}$,
the kinematic structure can be factored outside the integral
leaving a single integral expression for the
scattering amplitude.
As a direct consequence, the only dependence of the integrand on the
world-sheet coordinates comes from the exponential of the bosonic Green
function $G_{11}^{ij}$. However, in order that the amplitude be well
defined for non-trivial external momenta, we
must divide the integration region for the \mbox{Im\,$z_{i}$}
into three kinematic regions where each piece can be analytically
continued separately in the external momenta~\cite{me}.
In the zero-slope limit these
will give only the three distinct channels of 1PI graphs and no 1PR
graphs. We quote here for completeness the well
known result~\cite{schwarz}.
\beqa
\Gamma_{4}
&=& \frac{\kappa^{4}}{\al}
\left[ \prod_{i=1}^{4} \epsilon_{i}^{\mu_{i} \nu_{i}} \right]
K_{\mu_{1} \mu_{2} \mu_{3} \mu_{4}}
\ov{K}_{\nu_{1} \nu_{2} \nu_{3} \nu_{4}} \nonumber\\
& & \times \int_{{\cal F}} \frac{d^{2}\tau}{(\tau_{2})^{2}}
\int \left[ \prod_{i=1}^{3} \frac{d^{2} z_{i}}{\tau_{2}} \right]
\exp{ \left[ -\frac{i\pi \al}{2}
\sum_{ij}{}^{'} p_{i} \cdot p_{j} G_{11}^{ij} \right]} \nonumber\\
K_{\mu_{1} \mu_{2} \mu_{3} \mu_{4}}
&=& -\frac{1}{2} \left[ g_{\mu_{1} \mu_{3}} g_{\mu_{2} \mu_{4}} st
+ g_{\mu_{1} \mu_{4}} g_{\mu_{2} \mu_{3}} su
+ g_{\mu_{1} \mu_{2}} g_{\mu_{3} \mu_{4}} tu \right] \nonumber\\
& & + s\left[ g_{\mu_{1} \mu_{3}} p_{1\mu_{4}} p_{3\mu_{2}}
+ g_{\mu_{1} \mu_{4}} p_{1\mu_{3}} p_{4\mu_{2}}
+ g_{\mu_{2} \mu_{3}} p_{2\mu_{4}} p_{3\mu_{1}}
+ g_{\mu_{2} \mu_{4}} p_{2\mu_{3}} p_{4\mu_{1}} \right] \nonumber\\
& & + t\left[ g_{\mu_{1} \mu_{2}} p_{1\mu_{3}} p_{2\mu_{4}}
+ g_{\mu_{1} \mu_{3}} p_{1\mu_{2}} p_{3\mu_{4}}
+ g_{\mu_{2} \mu_{4}} p_{2\mu_{1}} p_{4\mu_{3}}
+ g_{\mu_{3} \mu_{4}} p_{3\mu_{1}} p_{4\mu_{2}} \right] \nonumber\\
& & + u\left[ g_{\mu_{1} \mu_{2}} p_{1\mu_{4}} p_{2\mu_{3}}
+ g_{\mu_{1} \mu_{4}} p_{1\mu_{2}} p_{4\mu_{3}}
+ g_{\mu_{2} \mu_{3}} p_{2\mu_{1}} p_{3\mu_{4}}
+ g_{\mu_{3} \mu_{4}} p_{3\mu_{2}} p_{4\mu_{1}} \right] \, ,
\nonumber\\
\eeqa
where $s$, $t$, and $u$ are the usual Mandelstam invariants,
$g_{\mu_{i}\mu_{j}}$ is the flat spacetime metric, and we
have used translational invariance to fix~$z_{4} = \tau$.
When written in this form the modular invariance of the
four-graviton amplitude is trivial to verify. The bosonic
Green function $G_{11}(z,\ov{z}| \tau, \ov{\tau})$ given in
eq.~(\ref{Gdef}) is invariant under the modular transformation
$\tau \rightarrow \frac{ a\tau + b}{c\tau +d}$ in conjunction with
the conformal change of coordinates
$z \rightarrow \frac{z}{c\tau + d}$. One recovers the original integraton
region for the world-sheet coordinates $z_{i}$ by making appropriate
lattice translations which are allowed due to the double periodicity
of the integrand.

The five-graviton amplitude is the highest order one-loop
graviton scattering amplitude for which the odd spin structure
contribution vanishes due to the integration over the Dirac
zero modes~\cite{D'Hoker}. Therefore, the full amplitude can
be obtained from eq.~(\ref{amp1}) by setting $N=5$ and performing
all the Grassman integrations. Since there are now five vertex
operators, we will necessarily generate terms
with products of five fermionic Green functions $S_{ab}^{ij}$
which must be summed over even spin structures.
In fact, a large number of such terms do appear
in addition to others coming from the various factors
in eq.~(\ref{eta}).
However, we now have no Riemann identity to simplify
the sum over spin structures and eliminate the $z_{i}$
dependence. Rather than presenting the complete result including
the numerous permutations of both the spacetime indices and
the arguments of the $S_{ab}^{ij}$ and $G_{11}^{ij}$,
we list only the various types
of terms which occur in order to verify the modular invariance
of the amplitude and examine its zero-slope limit.

After performing all the Grassman integrations, the final form
of the one-loop five-graviton amplitude can be written as
\beqa
\Gamma_{5}
&=& \kappa^{5}
\left[ \prod_{i=1}^{5} \epsilon_{i}^{\mu_{i} \nu_{i}} \right]
\int_{{\cal F}} \frac{d^{2}\tau}{\tau_{2}}
\int \left[ \prod_{i=1}^{4} \frac{d^{2} z_{i}}{\tau_{2}} \right]
\exp{ \left[ -\frac{i\pi \al}{2}
\sum_{ij}{}^{'} p_{i} \cdot p_{j} G_{11}^{ij} \right]} \nonumber\\
& & \times \int \left[ \prod_{\mu} dq^{\mu} \right]
e^{-\pi q^{2}} \,\,\,
{\cal K}_{\mu_{1} \mu_{2} \mu_{3} \mu_{4} \mu_{5}}
(z_{i}, \tau ; p_{i}, q) \,\,\,
\ov{\cal K}_{\nu_{1} \nu_{2} \nu_{3} \nu_{4} \nu_{5}}
(\ov{z}_{i}, \ov{\tau} ; p_{i}, q) \, ,
\nonumber\\
\label{amp5}
\eeqa
where we have rescaled $q^{\mu} \rightarrow
\sqrt{\frac{4\al}{\tau_{2}}} q^{\mu}$ and used translational
invariance to fix~$z_{5} = \tau$.
The only terms contributing to the
${\cal K}_{\mu_{1} \mu_{2} \mu_{3} \mu_{4} \mu_{5}}$
are those resulting from either products of four world-sheet fermionic
Green functions $S_{ab}^{ij}$ or five of them.
Each of the terms with four~$S_{ab}^{ij}$
multiplies a kinematic factor containing four powers
of the external momenta and can be simplified
with the identity~(\ref{riem}).
Let ${\cal P}_{\mu_{i_{1}} \mu_{i_{2}} \mu_{i_{3}} \mu_{i_{4}}}$
stand for the possible products of external momenta
$p_{i\mu_{j}}$, external
momentum invariants $p_{ij} = -(p_{i}+p_{j})^{2}$, and spacetime
metrics $g_{\mu_{i} \mu_{j}}$.
The terms coming from the products of four $S_{ab}^{ij}$
have the general form
\beq
{\cal P}_{\mu_{i_{1}} \mu_{i_{2}} \mu_{i_{3}} \mu_{i_{4}}}
\left[ \sum_{j \neq i_{5}}
\left(\partial_{i_{5}}G_{11}^{ji_{5}} \right) p_{j\mu_{i_{5}}}
+ \left( \frac{1}{\al \tau_{2}} \right)^{\frac{1}{2}}
q_{\mu_{i_{5}}} \right] \, ,
\label{gi5}
\eeq
where the indices
$(i_{1},i_{2},i_{3},i_{4},i_{5})$ are some permutation of
$(1,2,3,4,5)$. The sum over spin
structures has already been performed thereby eliminating the
dependence on the $S_{ab}^{ij}$.
The terms with products of five $S_{ab}^{ij}$ can be written
in the same notation where the sum over spin structures remains
explicit. In this case, however, ${\cal P}_{\mu_{i_{1}}
\mu_{i_{2}} \mu_{i_{3}} \mu_{i_{4}} \mu_{i_{5}}}$
will be fifth order in powers of the external momenta.
There are two types of terms.
\beq
{\cal P}_{\mu_{i_{1}} \mu_{i_{2}} \mu_{i_{3}} \mu_{i_{4}} \mu_{i_{5}}}
\sum_{a \times b = 0}(-)^{a+b}
\left\{ \frac{\Theta_{ab}(0|\tau)}{[\eta(\tau)]^{3}} \right\}^{4}
S_{ab}^{i_{1}i_{2}}S_{ab}^{i_{2}i_{3}}S_{ab}^{i_{3}i_{4}}
S_{ab}^{i_{4}i_{5}}S_{ab}^{i_{5}i_{1}} \, ,
\label{1PR1}
\eeq
and
\beq
{\cal P}_{\mu_{i_{1}} \mu_{i_{2}} \mu_{i_{3}} \mu_{i_{4}} \mu_{i_{5}}}
\sum_{a \times b = 0}(-)^{a+b}
\left\{ \frac{\Theta_{ab}(0|\tau)}{[\eta(\tau)]^{3}} \right\}^{4}
S_{ab}^{i_{1}i_{2}}S_{ab}^{i_{2}i_{1}}S_{ab}^{i_{3}i_{4}}
S_{ab}^{i_{4}i_{5}}S_{ab}^{i_{5}i_{3}} \, .
\label{1PR2}
\eeq

The modular invariance of eq.~(\ref{amp5}) can be checked easily by
noting that an even number of powers of loop momenta $q_{\mu}$ are needed
to give a non-zero contribution to the integral.
Since no cross terms present in
${\cal K}_{\mu_{1} \mu_{2} \mu_{3} \mu_{4} \mu_{5}}
\ov{\cal K}_{\nu_{1} \nu_{2} \nu_{3} \nu_{4} \nu_{5}}$
containing a single
power of $q_{\mu}$ appear, we get the necessary factor $1/\tau_{2}$
from the $q_{\mu_{i}}q_{\nu_{j}}$ piece. We will see in the next
section that this term gives 1PI Feynman graphs in the low-energy limit.
The properties of the world-sheet
Green functions under modular transformations are listed in the
appendix, and it is a simple matter to verify the modular invariance
of the remaining terms in eq.~(\ref{amp5}).

When written as a single integral expression,
the five-graviton amplitude~(\ref{amp5})
diverges for non-trivial values of the
external momenta for the same reason as the four-graviton
amplitude~\cite{amano}. Following~\cite{me}
we first re-write\footnote{Here we change the
normalization of the world-sheet bosonic Green function which is
allowed by momentum conservation and the graviton mass shell
condition.}
\beq
\exp{ \left[ -\frac{i\pi \al}{2}
\sum_{ij}{}^{'} p_{i} \cdot p_{j} G_{11}^{ij} \right]}
= \exp{\left[ \pi \al \tau_{2} \Phi \right]}
\prod_{1 \leq i<j \leq 5}
\tilde{\chi}_{ij}^{-\frac{1}{2} \al p_{ij}} \, ,
\eeq
where
\beqa
\Phi
&=& -\frac{1}{2} \sum_{1 \leq i<j \leq 5} p_{ij}
\left[ \frac{|y_{i}-y_{j}|}{\tau_{2}} \,-\,
\left(\frac{y_{i}-y_{j}}{\tau_{2}} \right)^{2} \right] \nonumber\\
\tilde{\chi}_{ij}
&=& \left| (1-e^{2i\pi z_{ij}}) \prod_{n=1}^{\infty}
\frac{(1 - e^{2i\pi \tau n + 2i\pi z_{ij}})
(1 - e^{2i\pi \tau n - 2i\pi z_{ij}})}
{(1 - e^{2i\pi \tau n})^{2}} \right| \, .
\nonumber\\
\eeqa
Next we divide the integration region for $y_{i}$ into the $4!/2$
independent orderings of the difference $y_{i}-y_{j}$.
The factor of $1/2$ comes from the discrete diffeomorphism
$z_{i} \rightarrow -z_{i} + \tau + 1$ as explained in ref.~\cite{me}.
For each of the twelve orderings there exists a convenient
set of five independent invariants $p_{ij}$ which simplify the
form of the function~$\Phi$. For each kinematic region, the
corresponding contribution must be analytically continued
independently in the external momenta in order to preserve
momentum conservation while on the graviton mass shell.
For example we find after
rescaling $y_{i} \rightarrow \tau_{2} y_{i}$
\beqa
\Phi(p_{12},p_{23},p_{34},p_{45},p_{51})
&=& p_{12}y_{1}(y_{3}-y_{2})
+ p_{23}(y_{4}-y_{3})(y_{2}-y_{1}) \nonumber\\
& & + p_{34}(1-y_{4})(y_{3}-y_{2})
+ p_{45}y_{1}(y_{4}-y_{3})  \nonumber\\
& & + p_{51}(1-y_{4})(y_{2}-y_{1}) \, ,
\label{phi}
\eeqa
for the ordering
$1 \leq y_{4} \leq y_{3} \leq y_{2} \leq y_{1} \leq 0$.
This contribution to the amplitude therefore converges for
${\rm Re}\,p_{12} \le 0$, ${\rm Re}\,p_{23} \le 0$,
${\rm Re}\,p_{34} \le 0$, \mbox{${\rm Re}\,p_{45} \le 0$}, and
${\rm Re}\,p_{51} \le 0$.

\section{The zero-slope limit}

The zero-slope limit of eq.~(\ref{amp5}) can now be taken in a
straightforward way. The most obvious source of 1PI graphs
comes from the terms proportional to the loop momenta~$q_{\mu}$
found in eq.~(\ref{gi5}).
As already mentioned, the terms linear in
$q_{\mu}$ do not contribute. The integral over
the quadratic terms $q_{\mu_{i}}q_{\nu_{j}}$ must be proportional
to the spacetime metric $g_{\mu_{i}\nu_{j}}$. For a particular
ordering of the~$y_{i}$ we get a kinematic factor times
\beqa
\Gamma_{5}^{({\rm 1PI})}
&=&
\frac{\kappa^{5}}{\al} \int_{\cal F} \frac{d^{2}\tau}{(\tau_{2})^{2}}
\int \left[ \prod_{i=1}^{4} d^{2}z_{i} \right]
\exp{ \left[\pi \al \tau_{2} \Phi \right]}
\prod_{ij}
\tilde{\chi}_{ij}^{-\frac{1}{2} \al p_{ij}} \, ,
\label{1PI}
\eeqa
where we have rescaled $y_{i} \rightarrow \tau_{2} y_{i}$, and the
product is understood to be ordered such that~$y_{i}-y_{j} \geq 0$.

To perform the zero-slope limit of eq.~(\ref{1PI}) we proceed as
in~\cite{me}. First divide
the fundamental domain into two regions with
\mbox{\({\cal F} = {\cal F}_{1} + {\cal F}_{2}\)}, where
\beqa
{\cal F}_{1} &:& -\frac{1}{2}\leq \tau_{1} \leq \frac{1}{2} \, ,
\;\;\; {\rm and} \;\;\;
\sqrt{1-(\tau_{1})^{2}} \leq \tau_{2} \leq 1 \, . \nonumber\\
{\cal F}_{2} &:& -\frac{1}{2}\leq \tau_{1} \leq \frac{1}{2} \, ,
\;\;\; {\rm and} \;\;\; 1 \leq \tau_{2} \leq \infty \, .
\eeqa
In region ${\cal F}_{1}$ we get
\beq
\left.
\lim_{\al \rightarrow 0} \Gamma_{5}^{({\rm 1PI})} \right|_{{\cal F}_{1}}
= \frac{\kappa^{5}}{\al} \left( \frac{\pi}{3} - 1 \right)
+ {\cal O}(\al) \, ,
\eeq
where the term zeroth order in $\al$ vanishes due to momentum
conservation and the graviton mass shell condition.
For region ${\cal F}_{2}$ we find after rescaling
$\tau_{2} \rightarrow \tau_{2}/(\pi \al)$
\beqa
\left.
\lim_{\al \rightarrow 0} \Gamma_{5}^{({\rm 1PI})} \right|_{{\cal F}_{2}}
&=& 2 \pi \kappa^{5}
\int_{\pi \al}^{\infty} \frac{d\tau_{2}}{(\tau_{2})^{2}}
\int_{0}^{1} \left[ \prod_{i=1}^{5} d\beta_{i} \right]
\delta \left( 1-\sum_{j=1}^{5}\beta_{j} \right) \nonumber\\
& & \times \left[
e^{\tau_{2} \Phi(p_{12},p_{23},p_{34},p_{45},p_{51})}
+ e^{\tau_{2} \Phi(p_{12},p_{23},p_{35},p_{54},p_{41})}
\right. \nonumber\\
& & + e^{\tau_{2} \Phi(p_{12},p_{24},p_{43},p_{35},p_{51})}
+ e^{\tau_{2} \Phi(p_{12},p_{24},p_{45},p_{53},p_{31})} \nonumber\\
& & + e^{\tau_{2} \Phi(p_{12},p_{25},p_{53},p_{34},p_{41})}
+ e^{\tau_{2} \Phi(p_{12},p_{25},p_{54},p_{43},p_{31})} \nonumber\\
& & + e^{\tau_{2} \Phi(p_{13},p_{32},p_{24},p_{45},p_{51})}
+ e^{\tau_{2} \Phi(p_{13},p_{32},p_{25},p_{54},p_{41})} \nonumber\\
& & + e^{\tau_{2} \Phi(p_{13},p_{34},p_{42},p_{25},p_{51})}
+ e^{\tau_{2} \Phi(p_{13},p_{35},p_{52},p_{24},p_{41})} \nonumber\\
& & \left.
+ e^{\tau_{2} \Phi(p_{14},p_{42},p_{23},p_{35},p_{51})}
+ e^{\tau_{2} \Phi(p_{14},p_{43},p_{32},p_{25},p_{51})} \right]
\,+\, {\cal O}(\al) \, . \nonumber\\
\label{zslfd}
\eeqa
We have introduced Feynman parameters $\beta_{i}$~\cite{'tHooft}
in place of the rescaled $y_{i}$ found in eq.~(\ref{phi}) in order
to make the field theory interpretation of the result more obvious.
\beq
\Phi(p_{12},p_{23},p_{34},p_{45},p_{51})
= p_{12}\beta_{1} \beta_{3}
+ p_{23}\beta_{2} \beta_{4}
+ p_{34}\beta_{3} \beta_{5}
+ p_{45}\beta_{4} \beta_{1}
+ p_{51}\beta_{5} \beta_{2}  \, .
\label{phiF}
\eeq
This is a sum of the twelve independent channels of pentagon Feynman
diagrams in the proper-time representation for a scalar theory
with cubic interactions.

To proceed with the zero-slope limit
we integrate by parts with respect to $\tau_{2}$ twice and use
momentum conservation and the graviton mass shell condition
to cancel the $\log{(\pi \al)}$ terms. The $\pi \al \log{(\pi \al)}$
terms cancel as a result of expanding the lower limit of the
$\tau_{2}$ integral. Therefore, to zeroth order in~$\al$ we get
\beqa
\left.
\lim_{\al \rightarrow 0} \Gamma_{5}^{({\rm 1PI})} \right|_{{\cal F}_{2}}
&=& \frac{\kappa^{5}}{\al} \,-\, 2\pi \kappa^{5}
\int_{0}^{1} \left[ \prod_{i=1}^{5} d\beta_{i} \right]
\delta \left( 1-\sum_{j=1}^{5}\beta_{j} \right)
\int_{0}^{\infty}d\tau_{2} \log{\tau_{2}} \nonumber\\
& & \times \left\{
\left[ \Phi(p_{12},p_{23},p_{34},p_{45},p_{51}) \right]^{2}
\exp{\left[\tau_{2}\Phi(p_{12},p_{23},p_{34},p_{45},p_{51}) \right]}
\right. \nonumber\\
& & +
\left[ \Phi(p_{12},p_{23},p_{35},p_{54},p_{41}) \right]^{2}
\exp{\left[ \tau_{2}\Phi(p_{12},p_{23},p_{35},p_{54},p_{41}) \right]}
\nonumber\\
& & +
\left[ \Phi(p_{12},p_{24},p_{43},p_{35},p_{51}) \right]^{2}
\exp{\left[ \tau_{2}\Phi(p_{12},p_{24},p_{43},p_{35},p_{51}) \right]}
\nonumber\\
& & +
\left[ \Phi(p_{12},p_{24},p_{45},p_{53},p_{31}) \right]^{2}
\exp{\left[ \tau_{2}\Phi(p_{12},p_{24},p_{45},p_{53},p_{31}) \right]}
\nonumber\\
& & +
\left[ \Phi(p_{12},p_{25},p_{53},p_{34},p_{41}) \right]^{2}
\exp{\left[ \tau_{2}\Phi(p_{12},p_{25},p_{53},p_{34},p_{41}) \right]}
\nonumber\\
& & +
\left[ \Phi(p_{12},p_{25},p_{54},p_{43},p_{31}) \right]^{2}
\exp{\left[ \tau_{2}\Phi(p_{12},p_{25},p_{54},p_{43},p_{31}) \right]}
\nonumber\\
& & +
\left[ \Phi(p_{13},p_{32},p_{24},p_{45},p_{51}) \right]^{2}
\exp{\left[ \tau_{2}\Phi(p_{13},p_{32},p_{24},p_{45},p_{51}) \right]}
\nonumber\\
& & +
\left[ \Phi(p_{13},p_{32},p_{25},p_{54},p_{41}) \right]^{2}
\exp{\left[ \tau_{2}\Phi(p_{13},p_{32},p_{25},p_{54},p_{41}) \right]}
\nonumber\\
& & +
\left[ \Phi(p_{13},p_{34},p_{42},p_{25},p_{51}) \right]^{2}
\exp{\left[ \tau_{2}\Phi(p_{13},p_{34},p_{42},p_{25},p_{51}) \right]}
\nonumber\\
& & +
\left[ \Phi(p_{13},p_{35},p_{52},p_{24},p_{41}) \right]^{2}
\exp{\left[ \tau_{2}\Phi(p_{13},p_{35},p_{52},p_{24},p_{41}) \right]}
\nonumber\\
& & +
\left[ \Phi(p_{14},p_{42},p_{23},p_{35},p_{51}) \right]^{2}
\exp{\left[ \tau_{2}\Phi(p_{14},p_{42},p_{23},p_{35},p_{51}) \right]}
\nonumber\\
& & + \left.
\left[ \Phi(p_{14},p_{43},p_{32},p_{25},p_{51}) \right]^{2}
\exp{\left[ \tau_{2}\Phi(p_{14},p_{43},p_{32},p_{25},p_{51}) \right]}
\right\} . \nonumber\\
\eeqa

As in the four-graviton amplitude, the contribution of the region
${\cal F}_{1}$ merely renormalizes the UV cut-off in the low-energy
field theory. We can replace $\pi \al \rightarrow 3\al$
in the lower limit of the $\tau_{2}$ integral in eq.~(\ref{zslfd})
to write the complete result as the proper-time representation of
1PI Feynman diagrams for a scalar theory with cubic interactions
with the modular parameter $\tau_{2}$ playing the role of the
proper-time variable.

The surprising feature of eq.~(\ref{zslfd}) is the power of $\tau_{2}$
occurring in the proper-time integral.
For a scalar $\phi^{3}$ theory in ten
dimensions with five external particles, one would find only a
single power of $\tau_{2}$ in the denominator of the
integrand (see appendix of~\cite{me}).
Instead we find the same UV degree of divergence as in the zero-slope
limit of the four-graviton amplitude. The form of the
function~$\Phi$ found in eq.~(\ref{phiF}) shows that
these 1PI Feynman graphs have
five internal propagators in a theory with cubic interactions.
However, the low-energy supergravity theory contains both cubic and
higher order interactions with non-trivial derivative couplings.
The field theory limit of superstring theory gives the result of
summing all of the contributions to the one-loop five-point
amplitude with the Lorentz structure already factored outside
the integrals over the Feynman parameters. Complicated cancellations
among diagrams must occur preserving gauge invariance which
produce a term having a single transverse kinematic factor
times the scalar integral found in eq.~(\ref{zslfd}).
Therefore, from the field theory point of view the UV degree of
divergence found in this expression results directly from the
underlying  gauge invariance of the
ten dimensional low-energy field theory.

{}From the string theory point of view the power of $\tau_{2}$
occurring in the integrand of eq.~(\ref{zslfd}) is required to preserve
the modular invariance of the superstring scattering amplitude.
The power of $\tau_{2}$
and therefore the degree of UV divergence will be the same no
matter how many external gravitons are inserted. One can prove
that such a term does come from the even spin structure
contribution to the N-graviton amplitude
given in eq.~(\ref{amp1}) as follows.
Initially there are six
powers of $\tau_{2}$ in the denominator before performing the
Grassman integrations. These 1PI graphs come from the terms
with unexponentiated powers of the loop momenta $q_{\mu_{i}}$
multiplying the  the $\exp{[-\frac{\pi \tau_{2}}{4\al} q^{2}]}$
factor. Only an even
number of such powers will contribute to the integral over the
loop momenta, and each pair of them gives a factor $1/ \tau_{2}$.
Each power of $q_{\mu_{i}}$ multiplies only a single
$\ov{\theta}_{i}$ and the sum over even spin structures requires
exactly four factors of the $S_{ab}^{ij}$ in order to eliminate
all of the $z_{i}$ dependence of the integrand
not found in the factor
\beq
\exp{\left[ -\frac{i\pi \al}{2} \sum_{ij}{}' p_{i} \cdot p_{j}
G_{11}^{ij} \right]} \, .
\eeq
Since we can always have four factors of the~$S_{ab}^{ij}$
each multiplying a single~$\theta_{i}$,
terms with $N-4$ powers of $q_{\mu_{i}}$ will occur in each
of the left and right moving sectors.
Overall, we will therefore get $6+N-4 = N+2$ powers of $\tau_{2}$
in the denominator for $N$ vertex insertions.
Each modular invariant integration measure
for the location of the vertex operator must get a $1/ \tau_{2}$
leaving exactly a $1/ (\tau_{2})^{2}$ for the integral
over~$d^{2}\tau$.

The rest of the diagrams appearing in the zero-slope limit
of the five-graviton amplitude~(\ref{amp5}) are generated as
follows. Using the representations for the chiral Green functions
found in the appendix, one finds after
rescaling $y_{i} \rightarrow \tau_{2} y_{i}$ and then
$\tau_{2} \rightarrow \tau_{2} / (\pi \al)$ that their
contribution to the zero-slope limit is given by
\beqa
\lim_{\al \rightarrow 0} \partial_{i} G_{11}^{ji}
&=& -(y_{i} - y_{j}) + {\cal O}(\al) \nonumber\\
\lim_{\al \rightarrow 0} S_{01}^{ij}
&=& {\cal O}(\al) \nonumber\\
\lim_{\al \rightarrow 0} S_{00}^{ij}
&=& {\cal O}(\al) \nonumber\\
\lim_{\al \rightarrow 0} S_{10}^{ij}
&=& \half + {\cal O}(\al) \, .
\eeqa
Since the $\tau_{2}$ integral diverges logarithmically in
the cut-off~$\al$, the terms of order~$\al$ in the above
expansions do not contribute to the zero-slope limit
for~$z_{i} \neq z_{j}$.
One must be carefull, however, since these functions
each have a single pole in $z_{i}-z_{j}$ with residue~$-1/2i\pi$.
The poles do contribute to the zero-slope limit as
$z_{i} \rightarrow z_{j}$. We therefore find a
second set of 1PI diagrams that comes from the $(y_{i}-y_{j})$ in
$\partial_{i} G_{11}^{ji}$ and the $1/2$ in~$S_{10}^{ij}$.
This set of 1PI graphs has the expected logarithmic
dependence on the cut-off but contains unexponentiated
Feynman parameters in the integrand corresponding to
derivative couplings in the low-energy field theory.
The zero-slope limit of the terms that
generate these graphs can be taken
as before, and we do not repeat the procedure here.

The 1PR graphs which appear in the zero-slope limit of the
five-graviton amplitude come from the absolute value squared of the
single poles in~$z_{i}-z_{j}$.
For these pole terms we cannot set~$\al$ to zero in the factor
$\prod_{ij} \tilde{\chi}_{ij}^{-\frac{1}{2} \al p_{ij}}$
without introducing a logarithmic divergence in the integration
over the coordinates of one of the vertex operators.
No cross terms between different
single poles contribute as~$\al \rightarrow 0$ since these do not
introduce any singularities and therefore vanish.
We must integrate the un-rescaled
$z_{i}$ vertex coordinate over a small disk about $z_{j}$
in order to pick up the pole in $\al p_{ij}$ and then take the zero-slope
limit as before. For example, consider the case where
$z_{i_{2}} \rightarrow z_{i_{1}}$ for some $(i_{1},i_{2})$.
In practice, we can formally include all of the twelve orderings
at once. We fix $z_{i_{1}}$ to be
anywhere on the torus due to translational invariance
(instead of $z_{5} = \tau$)
and integrate $z_{i_{2}}$ over a disk of radius
$\epsilon$ about that point. For the case of a product of five
fermionic Green functions, when one of them is near the pole
the sum over spin structure to eliminate the other four
can be performed using eq.~(\ref{riem}).
In order to include all possible channels in the zero-slope limit
each contribution to the pole must be considered separately.

The double
pole introduced by eq.~(\ref{1PR2}) will not contribute since
the sum over spin structures with only three propagators vanishes.
The integral over the phase angle ensures that only a $1/ r^{2}$
will survive from the product
${\cal K}_{\mu_{1} \mu_{2} \mu_{3} \mu_{4} \mu_{5}}
\ov{\cal K}_{\nu_{1} \nu_{2} \nu_{3} \nu_{4} \nu_{5}}$
in eq.~(\ref{amp5}) where $r = |z_{i_{2}}-z_{i_{1}}|$.
We then find up to
an overall kinematic factor tenth order in the external momenta
\beqa
\lim_{\al \rightarrow 0} \Gamma_{5}^{({\rm 1PR})}
&=& \kappa^{5}
\int_{{\cal F}} \frac{d^{2}\tau}{\tau_{2}}
\int \left[ \frac{d^{2} z_{i_{3}}}{\tau_{2}}
\frac{d^{2} z_{i_{4}}}{\tau_{2}}
\frac{d^{2} z_{i_{5}}}{\tau_{2}} \right] \nonumber\\
& & \times
\frac{1}{\tau_{2}} \int_{0}^{2\pi} d\phi
\int_{0}^{\epsilon} dr \,\, r^{-1}
\exp{ \left\{
-i\pi \al p_{i_{1}} \cdot p_{i_{2}} G_{11}^{i_{1}i_{2}}
\right\} } \nonumber\\
& & \times
\exp{ \left\{ -i\pi \al
\left[ p_{i_{5}} \cdot p_{i_{4}}G_{11}^{i_{5}i_{4}}
+ p_{i_{5}} \cdot p_{i_{3}}G_{11}^{i_{5}i_{3}}
+ p_{i_{4}} \cdot p_{i_{3}}G_{11}^{i_{4}i_{3}} \right. \right. }
\nonumber\\
& & \left. \left.
+ p_{i_{1}} \cdot \sum_{n=3}^{5} p_{i_{n}}G_{11}(z_{i_{n}})
+ p_{i_{2}} \cdot \sum_{n=3}^{5} p_{i_{n}}G_{11}(z_{i_{n}}-re^{i\phi})
\right] \right\} \,+\, \ldots
\nonumber\\
\eeqa
We now expand the rest of the integrand in $r$
keeping only the lowest order term. The bosonic Green function satisfies
$G_{11}^{i_{1}i_{2}} = -(1/ i\pi)\log{r} + \ldots$
and we get
\beqa
\lim_{\al \rightarrow 0} \Gamma_{5}^{({\rm 1PR})}
&=& 2\pi \kappa^{5} \left[ -\frac{2}{\al p_{i_{1}i_{2}}} \right]
\int_{{\cal F}} \frac{d^{2}\tau}{(\tau_{2})^{2}}
\int \left[ \frac{d^{2} z_{i_{3}}}{\tau_{2}}
\frac{d^{2} z_{i_{4}}}{\tau_{2}}
\frac{d^{2} z_{i_{5}}}{\tau_{2}} \right] \nonumber\\
& & \times
\exp{ \left\{ -i\pi \al
\left[ p_{i_{5}} \cdot p_{i_{4}}G_{11}^{i_{5}i_{4}}
+ p_{i_{5}} \cdot p_{i_{3}}G_{11}^{i_{5}i_{3}}
+ p_{i_{4}} \cdot p_{i_{3}}G_{11}^{i_{4}i_{3}} \right. \right. }
\nonumber\\
& & \left. \left.
+ \left(p_{i_{1}} + p_{i_{2}} \right)
\cdot \sum_{n=3}^{5} p_{i_{n}}G_{11}(z_{i_{n}})
\right] \right\} \,+\, \ldots
\eeqa

The above expression contains a massless pole
times the scalar part of the
four graviton amplitude with one external line having momenta
\mbox{$(p_{i_{1}} + p_{i_{2}})$}. When all of the pieces of the
five-graviton amplitude
are included, one would find all possible factorized
channels consisting of a tree level three-graviton amplitude
(just a kinematic factor) times the one-loop four-graviton amplitude.
The rest of the zero-slope limit can be performed exactly
as in ref.~\cite{me} giving 1PR Feynman graphs with a massless
single pole.

\section{Conclusion}

We have presented a concise and explicit calculation of the even
spin structure contribution to the one-loop N-graviton scattering
amplitude in the type-II superstring theory. In this calculation
we chose to use the path integral formulation of the NSR string
since this approach leads directly to a covariant result expressed
in terms of Jacobi theta functions whose properties under modular
transformations are well known. Limited superspace formalism was
used in the calculation since there is no topological
obstruction to the superconformal gauge for the one-loop
even spin structure contribution.

For $N \leq 5$ the even spin structure contribution
gives the complete amplitude, and the amplitudes for $N \leq 3$
vanish due to a Riemann identity for theta functions. The
$N=4$ result is easily reproduced, and its modular invariance is
trivial to verify. It must be separated into three terms, each of
which is convergent in a different kinematic region.
For $N=5$ we find in direct analogy that the final
result must be divided into twelve kinematic regions where
each piece can be analytically continued separately in the
external momenta. Chiral world-sheet Green functions present in
the integrand of the five-graviton amplitude produce massless
single-particle poles in the limit when the coordinates of two
vertex operators come close together.

In sect.~4
we examined the form of the zero-slope limit
of the one-loop five-graviton
amplitude and identified three types of contributions.
The first came from terms with two powers of loop momenta
$q_{\mu_{i}} q_{\nu_{j}}$ which are proportional to the
spacetime metric $g_{\mu_{i} \nu_{j}}$ after integration. All
of these terms have the form of a kinematic factor times the
scalar integral found in eq.~(\ref{1PI}).
As $\al \rightarrow 0$ these terms generate pentagon Feynman diagrams
in the proper-time representation for a scalar theory with cubic
interactions. The square
of the proper-time variable~$\tau_{2}$ occuring in the denominator
of the integrand gives the same degree of divergence in the UV limit
as that found in the 1PI graphs appearing in the zero-slope limit
of the four-graviton amplitude.
We proved that the modular invariance of the even spin structure
contribution to the N-graviton amplitude
guarantees that this set of 1PI graphs will have the same UV
degree of divergence for all~$N$. For $N > 5$, however, there
will also be a parity violating contribution coming from the odd spin
structure. It does not mix with the even spin structure pieces under
modular transformations, and therefore could have some different
dependence on~$\tau_{2}$ in the zero-slope limit.

The second contribution to the zero-slope limit of the
five-graviton amplitude comes from the non-singular part
of~$\partial_{i} G_{11}^{ji}$ and~$S_{10}^{ij}$.
These terms are proportional to the same scalar integral
found in eq.~(\ref{1PI}) but with only a single power of $\tau_{2}$
in the denominator of the integrand. They also give pentagon
diagrams and have the expected logarithmic divergence
in the cut-off~$\al$.
Some of the graphs also contain unexponentiated powers of the
Feynman parameters in the integrand corresponding to derivative
couplings in the low-energy field theory.

The third contribution to the low-energy limit for $N=5$
comes from the single poles in the chiral Green functions
as~$z_{i} \rightarrow z_{j}$. By integating over a small region
about these poles we obtain
1PR Feynman diagrams with a massless single-particle
pole in the zero-slope limit. The residue of each pole is
a kinematic factor times a sum of four-particle
box graphs. Although no 1PR graphs are present
in the zero-slope limit of the one-loop four-graviton amplitude
due to a simplification of its integrand resulting from a Riemann
identity for theta functions, we expect that they will contribute
for all~$N \ge 5$. Since each momentum invariant must multiply an~$\al$
on dimensional grounds, one cannot take the zero-slope limit
without introducing a single-particle pole as if
the sum of two external momenta are on the
graviton mass shell. Therefore, if massless poles exist in the
full superstring amplitude, then 1PR Feynman diagrams must contribute
directly in the zero-slope limit and are not generated by duality.
\\
\\
\noindent
I would like to thank W.I. Weisberger for useful discussions.

\appendix
\section{Appendix}

Here we state some properties of the world-sheet Green functions
and the Jacobi theta functions from which they are defined.
The world-sheet fermions $\psi (z)$ carry a spin structure
$[\nu_{a}, \nu_{b}]$ which can take on values $[\frac{1}{2}, \frac{1}{2}]$,
$[\frac{1}{2}, 0]$, $[0, \frac{1}{2}]$, and $[0,0]$. Their periodicity
on the torus is defined by
\beqa
\psi (z+1) &=& e^{2i\pi (\nu_{a} - \frac{1}{2})} \psi (z) \nonumber\\
\psi (z+\tau) &=& e^{2i\pi (\nu_{b} - \frac{1}{2})} \psi (z) \, .
\eeqa
Theta functions with characteristics corresponding to these spin
structures are defined by
\beq
\Theta_{[\nu_{a}, \nu_{b}]} (z| \tau)
= \sum_{n = -\infty}^{\infty}
\exp{ \left\{ 2i\pi \left[ \frac{1}{2}(n+\nu_{a})^{2} \tau
+ (n+\nu_{a})(z+\nu_{b}) \right] \right\} } \, .
\eeq
We also have the Dedekind eta function given by
\beq
\eta (\tau) = \exp{[\frac{i\pi \tau}{12}]}
\prod_{n=1}^{\infty} \left( 1 - e^{2i \pi \tau n} \right) \, .
\eeq

Superstring scattering amplitudes are invariant under the modular
transformations $\tau \rightarrow \frac{a\tau + b}{c\tau + d}$
in conjunction with the conformal change of coordinates
$z \rightarrow \frac{z}{c\tau + d}$, where $a,b,c,d$ are integers.
These transformations are generated by $\tau \rightarrow \tau + 1$
and $\tau \rightarrow -1/ \tau$. For the properties under these
transformations of the Dedekind eta function and the Jacobi theta
functions from which the world-sheet Green functions are defined
see ref.~\cite{mumford}.
The bosonic world-sheet Green function $G_{11}(z|\tau)$
given in eq.~(\ref{Gdef}) can then be shown to satisfy
$G_{11}(\frac{z}{c\tau + d} | \frac{a\tau + b}{c\tau + d})
= G_{11}(z|\tau)$. The even spin structure fermionic Green functions
$S_{ab}(z|\tau)$ satisfy
\beqa
S_{10}(z| \tau + 1) &=& S_{10}(z| \tau) \nonumber\\
S_{00}(z| \tau + 1) &=& S_{01}(z| \tau) \nonumber\\
S_{01}(z| \tau + 1) &=& S_{00}(z| \tau) \, ,
\eeqa
and
\beqa
S_{10} \left( \frac{z}{\tau} | -\frac{1}{\tau} \right)
&=& \tau S_{01}(z| \tau) \nonumber\\
S_{00} \left( \frac{z}{\tau} | -\frac{1}{\tau} \right)
&=& \tau S_{00}(z| \tau) \nonumber\\
S_{01} \left( \frac{z}{\tau} | -\frac{1}{\tau} \right)
&=& \tau S_{10}(z| \tau) \, .
\eeqa

One can prove a form of the Riemann identity for theta functions
which can be written
\beq
\sum_{\nu} (-)^{a+b} \prod_{i=1}^{4} \Theta_{ab} (z_{i}| \tau)
= 2 \prod_{i=1}^{4} \Theta_{11} (w_{i}| \tau) \, ,
\eeq
where
\beqa
w_{1} &=& \frac{1}{2} (z_{1} + z_{2} + z_{3} + z_{4}) \nonumber\\
w_{2} &=& \frac{1}{2} (z_{1} + z_{2} - z_{3} - z_{4}) \nonumber\\
w_{3} &=& \frac{1}{2} (z_{1} - z_{2} + z_{3} - z_{4}) \nonumber\\
w_{4} &=& \frac{1}{2} (z_{1} - z_{2} - z_{3} + z_{4}) \, .
\eeqa
Using the definition of $S_{ab}^{ij}$ given in eq.~(\ref{Gdef})
and the fact that $\Theta_{11}(0|\tau) = 0$ one can then show
\beqa
\sum_{a \times b = 0}(-)^{a+b}
\left\{ \frac{\Theta_{ab}(0| \tau)}{[\eta (\tau)]^{3}} \right\}^{4}
S_{ab}^{i_{1}i_{2}} S_{ab}^{i_{2}i_{1}} &=& 0 \nonumber\\
\sum_{a \times b = 0}(-)^{a+b}
\left\{ \frac{\Theta_{ab}(0| \tau)}{[\eta (\tau)]^{3}} \right\}^{4}
S_{ab}^{i_{1}i_{2}} S_{ab}^{i_{2}i_{3}}
S_{ab}^{i_{3}i_{1}} &=& 0 \nonumber\\
\sum_{a \times b = 0}(-)^{a+b}
\left\{ \frac{\Theta_{ab}(0| \tau)}{[\eta (\tau)]^{3}} \right\}^{4}
S_{ab}^{i_{1}i_{2}}S_{ab}^{i_{2}i_{3}}
S_{ab}^{i_{3}i_{4}}S_{ab}^{i_{4}i_{1}}
&=& -1 \, .
\eeqa

We may obtain representations of the chiral Green functions as
infinite sums which are useful
for calculating the low-energy limit of superstring scattering
amplitudes. We begin with the mode expansion for the Green function
and then perform one of the two sums using the Sommerfeld-Watson
transformation.
\beqa
\partial \left[G_{11} + \frac{1}{4i\tau_{2}} (z-\ov{z})^{2} \right]
&=& \half + \frac{ e^{2i\pi z}}{1 - e^{2i\pi z}} \nonumber\\
& & + \sum_{m=1}^{\infty} \left[
\frac{ e^{2i\pi m (\tau + z)} - e^{2i\pi m (\tau - z)}}
{1 - e^{2i\pi m \tau}} \right] \, ,
\eeqa
valid for $0 \le {\rm Im}\, z \le {\rm Im}\, \tau$.
\beq
S_{01}(z) = \frac{ e^{i\pi z}}{1 - e^{2i\pi z}}
+ \sum_{m=1}^{\infty} \left[
\frac{ e^{2i\pi (m - 1/2)(\tau + z)} - e^{2i\pi (m - 1/2)(\tau - z)}}
{1 - e^{2i\pi (m - 1/2) \tau}} \right] \, ,
\eeq
valid for $0 \le {\rm Im}\, z \le {\rm Im}\, \tau$.
\beq
S_{00}(z) = \frac{ e^{i\pi z}}{1 - e^{2i\pi z}}
- \sum_{m=1}^{\infty} \left[
\frac{ e^{2i\pi (m - 1/2)(\tau + z)} - e^{2i\pi (m - 1/2)(\tau - z)}}
{1 + e^{2i\pi (m - 1/2) \tau}} \right] \, ,
\eeq
valid for $0 \le {\rm Im}\, z \le {\rm Im}\, \tau$.
\beq
S_{10}(z) = \frac{1}{\tau}
\frac{ e^{i\pi z/ \tau}}{1 - e^{2i\pi z/ \tau}}
- \frac{1}{\tau} \sum_{m=1}^{\infty} \left[
\frac{ e^{2i\pi (m - 1/2) z / \tau}
- e^{-2i\pi (m - 1/2) z / \tau}}
{1 - e^{2i\pi (m - 1/2)/ \tau}} \right] \, ,
\eeq
valid for $\frac{\tau_{1}}{\tau_{2}} {\rm Im}\, z
\le {\rm Re}\, z \le \frac{\tau_{1}}{\tau_{2}} {\rm Im}\, z + 1$.

\end{document}